\documentclass[aps,prl,groupedaddress,reprint,showpacs,pdftex]{revtex4-1}
\usepackage{graphicx}
\usepackage{tabularx}
\usepackage{bm}
\usepackage{amsmath}
\usepackage{amssymb}
\usepackage{txfonts}
\usepackage{url}

\usepackage[usenames,dvipsnames]{color}
\definecolor{Gray}{gray}{0.0}
\definecolor{lightGray}{gray}{0.35}
\begin{document}
\title{
  A Quantum Annealing Approach to Ionic Diffusion in Solids
}
\author{Keishu Utimula$^{1}$}
\email{mwkumk1702@icloud.com}
\author{Tom Ichibha$^{2}$}
\author{Genki I. Prayogo$^{1}$}
\author{Kenta Hongo$^{3,4,5}$}
\author{Kousuke Nakano$^{2}$}
\author{Ryo Maezono$^{2}$}

\affiliation{$^{1}$
  School of Materials Science, JAIST, Asahidai 1-1, Nomi, Ishikawa,
  923-1292, Japan
}
\affiliation{$^{2}$
  School of Information Science, JAIST, Asahidai 1-1, Nomi, Ishikawa,
  923-1292, Japan
}
\affiliation{$^{3}$
  Research Center for Advanced Computing Infrastructure,
  JAIST, Asahidai 1-1, Nomi, Ishikawa 923-1292, Japan
}
\affiliation{$^{4}$
  Center for Materials Research by Information Integration,
  Research and Services Division of Materials Data and
  Integrated System, National Institute for Materials Science,
  Tsukuba 305-0047, Japan
}
\affiliation{$^{5}$
  PRESTO, Japan Science and Technology Agency, 4-1-8 Honcho,
  Kawaguchi-shi, Saitama 322-0012, Japan
}

\date{\today}
\begin{abstract}
  We have developed a framework for using quantum annealing
computation to evaluate a key quantity in 
ionic diffusion in solids, the correlation factor. 
Existing methods can only calculate the
correlation factor analytically in the case of physically 
unrealistic models, making it difficult to relate microstructural 
information about diffusion path networks obtainable 
by current {\it ab initio} techniques to macroscopic 
quantities such as diffusion coefficients.
We have mapped the problem into 
a quantum spin system described by the Ising Hamiltonian. 
By applying our framework in combination
with ab initio technique, 
it is possible to understand how 
diffusion coefficients are controlled by 
temperatures, pressures, atomic substitutions, and other  
factors.
We have calculated the correlation factor
in a simple case with a known exact result by a variety of
computational methods, including simulated quantum annealing on the spin models, 
the classical random walk, the matrix description,
  and quantum annealing on D-Wave with hybrid solver
. 
This comparison shows that 
all the evaluations give 
consistent results with each other, but that many
of the conventional approaches require infeasible 
computational costs. Quantum annealing is also currently infeasible 
because of the cost and scarcity of Q-bits, but we argue that when 
technological advances alter this situation, quantum 
annealing will easily outperform all existing methods.
\end{abstract}
\maketitle

\section{Introduction}
\label{sec.intro}
The quantum annealing technique~\cite{1998KAD, 2018KUM} has
been widely and successfully applied to
challenging combinatorial optimizations~\cite{2006SAN},
including NP(Non-deterministic Polynomial time)-hard and NP-complete problems~\cite{2014LUC, 2008DAS, 2001FAR, 2006SAN}. Realistic problems such as
the vehicle routing problem (CVRP),
optimization of traffic quantity~\cite{2017NEU, 2017SYR, 2013CRI, 2004MAR}, 
investment portfolio design~\cite{2015ROS},
scheduling problems~\cite{2015DAV}, and
digital marketing~\cite{2017TAK} have 
recently been addressed by quantum annealing.

\vspace{2mm}
In the chemistry and materials science domain, however, 
relatively few applications have been found, 
other than investigation of the molecular similarity 
problem~\cite{2017HER} or the search 
for protein conformations~\cite{2012PER}. 
This contrasts with the many applications 
of quantum gate computing to this 
field~\cite{2019CAO}, e.g., in quantum phase estimation. 
This imbalance is self-perpetuating:
chemists and materials scientists are
unfamiliar with quantum annealing,
and so do not think to use it.
Finding additional applications
of the technique is therefore important not
only for the sake of the applications themselves,
but also for the sake of increasing
recognition of quantum annealing
as a useful method in this domain. 

\vspace{2mm}
In the quantum annealing framework, 
an optimization problem 
is mapped into 
a quantum spin system 
described by the Ising Hamiltonian~\cite{1998KAD, 2018KUM}. 
The problem is then solved 
by searching for optimal spin 
configurations minimizing the 
energy of the Hamiltonian. 
In this framework,
the problem of finding an optimum
in the presence of many local minima
is solved by using quantum tunneling
(i.e. virtual hopping)
to cross high energy barriers.
The
quantum framework is an increasingly
popular tool for the solution of
optimization problems in the everyday, classical world. 
However, its application   
to problems 
in the quantum world~\cite{2017HER}
seems to be surprisingly rare. 
In the present study, we applied it to 
ionic diffusion in solids ~\cite{2007MEH}. 
This quantum-mechanical topic, which is of
great interest in both pure and applied materials 
science, originally attracted attention in connection with
the microscopic analysis of mechanical strengths
~\cite{2011KUM},
and more recently has been connected to the efficiency of batteries, 
systems where charge-carrying ions diffusing 
in the solid electrolyte are clearly of central importance 
~\cite{2013SHI, 2016BAC, 2009LEV}.

\vspace{2mm}
Among the various mechanisms~\cite{2007MEH} of 
ionic diffusion, 
we concentrate here on 
the vacancy mechanism ~\cite{2007MEH},
in which ions hop only 
between lattice sites.
Although many {\it ab initio} works 
have provided 
insight into {\it microscopically} 
'easier paths' for the ion to hop along, 
it remains difficult to get 
practically useful knowledge 
of the diffusion coefficient $D$ 
as a {\it macroscopic} quantity. 
To connect the microscopic knowledge 
with the macroscopic quantity, 
we must cope with the difficult problem of 
counting all possible processes 
by which an ion is pulled back 
toward a vacancy~\cite{2019ICHa}
(while also being 
pulled in other directions, 
as explained in the next section). 
This process is described by the {\it correlation factor}
~\cite{2007MEH, 2019ICHa}
$f$.
The evaluation of $f$, which involves identifying the optimum routing 
as a vacancy hops around on lattice sites 
for a given anisotropic {\it easiness},
is essential for 
connecting 
the microscopic analysis with 
the evaluation of practically useful 
macroscopic quantities ~\cite{2019ICHa}. 
Such a routing problem is analogous 
to classical ones that have been 
successfully treated in the annealing framework. 
Otherwise, the evaluation is far too difficult 
to solve in the general case;  
so far, only very limited cases and simple models
({e.g.}, the simple cubic lattice) have been solved ~\cite{2007MEH}. 
In the present work, 
we provide a way to formulate the evaluation 
in the annealing framework, and 
show that the method successfully
overcomes difficulties 
unsolved by conventional approaches. 

\section{Formulation}
\subsection{Correlation factor in diffusion mechanism}
We consider a form of atomic diffusion where 
the atom to be considered (the 'tracer') hops onto 
a neighboring vacancy site ('hole') 
generated by 
thermal processes. 
Let the tracer be
located on a site $\alpha$.
At the initial step ($i=0$),
we will write $\alpha = S$ (\underline {S}tart).
Any hopping of the tracer onto 
neighboring vacant sites generates a 
hole on $\alpha = S$ at the $i=1$ step. 
This hole then becomes a possible 
vacant site by which the tracer may get back 
to $\alpha = S$, a process described as   
'the hole pulls the tracer back with a 
certain probability'. 
This probability is typically manifest as a 
reduction of the effective 
stride of the tracer by a factor $f$,
the 
correlation factor of the 
diffusion.   
\begin{figure}[htbp]
  \includegraphics[width=\hsize]{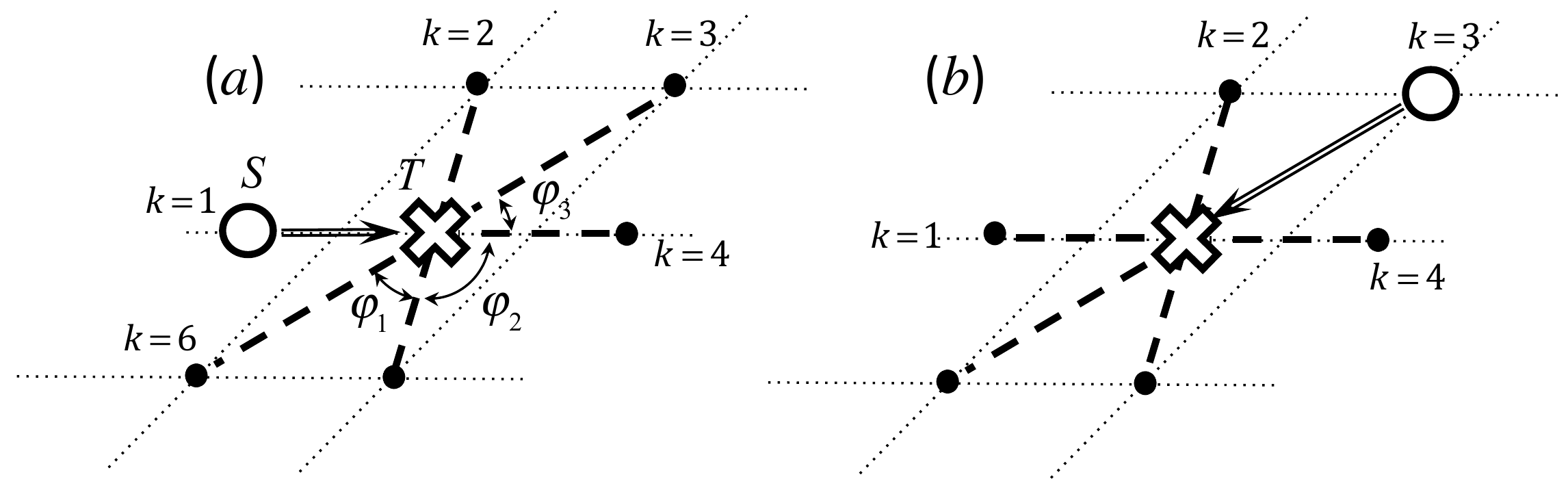}
  \caption{\label{theta}
      Examples of snapshots for the vacancy
      (white circles, initially at site $S$)
      to attract a tracer (white crosses at site $T$)
      to the vacancy's position.
      The horizontal direction to the right
      is defined to be identical to that of the diffusion flow
      to be considered. 
      The vacancy is located at one of the $Z$ neighboring sites 
      to site $T$ ($Z$=6 as an example in the panels) right before
      exchanging positions with the tracer.
      The vacancy site is denoted by $k$. 
      The attraction angles from site $k$ 
      are $\theta_1 = \pi$, $\theta_2 = \pi - \varphi_2$, 
      $\theta_3 =  \varphi_3$, $\theta_4 = 0$, $\cdots$.
      The panel (a) indicates the most likely case that
      the vacancy pulls behind the tracer and the panel (b)
      indicates that the vacancy pulls forward the tracer
      after detour movements.
  }
\label{theta}
\end{figure}


\vspace{2mm}
While the simplest picture would be 
an immediate 'pull-back' made by a vacancy 
at $\alpha = S$ when $i=2$, 
we must take into account further ways 
a wandering vacancy can attract a tracer 
when $i\ge 3$. 
We shall therefore consider the final state 
(where the vacancy is about to attract a tracer). 
Let the site $\alpha = T$ be where the tracer 
is located at step $i=(N-1)$, immediately before it is finally attracted back to the neighboring vacancy. Because this is an exchange process,
the vacancy will be located at $\alpha = T$ when $i=N$. 
To specify the geometry, let $\theta =0$ be 
the direction of the diffusion flow with a 
radius vector centered at $\alpha = T$
~(Fig.~{\ref{theta}}). 
Let the number of neighboring sites to $\alpha = T$ 
be $Z$, with locations specified by $\theta_k$. 
A
  pulling back
by a vacancy at $\theta_k$ 
is then contributing to the diffusion by its projection, 
$\cos\theta_k$. 
Letting $P_k$ be the probability distribution to 
get a vacancy at a specific $\theta_k$ amongst $Z$ 
when $i=(N-1)$, 
  the 'average cosine'
, 
\begin{eqnarray}
  \left\langle {\cos \theta } \right\rangle
    = \sum\limits_{k=1}^Z {{P_k}\cos {\theta _k}} \ , 
    \label{eq.average_cosine}
\end{eqnarray}
matters to the correlation factor. 
Further consideration is required 
to take into account the fact that 
a pulling-back process itself is also 
subject to pulling-back. 
Such multiple processes are finally 
convoluted into a form
~\cite{2019ICHa,1956COM} as, 
\begin{eqnarray}
  f &=& 1 + 2\sum\limits_{n = 1}^\infty  
        {\left\langle {\cos \theta } \right\rangle ^n}
        =
        \frac{
          {1+\left\langle {\cos \theta }
            \right\rangle}
        }{1-{\left\langle {\cos \theta }
            \right\rangle}} \ .
        \label{eq.correlation_factor}
\end{eqnarray}
With $\theta$ 
as in Fig.~\ref{theta}, 
this factor ranges from 
$f = 0$ ($\theta = \pi$) 
through $f = 1$ ($\theta = \pi /2$) 
to $f\to \infty$ ($\theta\to 0$). 

\subsection{Formulation using quantum annealing Hamiltonian}
The evaluation of the correlation factor is 
therefore reduced to the calculation of the 
averaged projection given in Eq.(\ref{eq.average_cosine}). 
The mission of the simulations is to 
provide the probability $P_k$, which is 
obtained from 
the trajectories of a vacancy hopping 
along optimal paths in the given system,
i.e., 
those 
satisfying the initial [$\alpha = S$ ($i=1$)] 
and the final [$\alpha = T$ ($i=N$)] 
conditions: 
the probability distribution for 
these trajectories gives $P_k$ at $i=(N-1)$. 

\vspace{2mm}
The problem of getting the optimum trajectories 
is well formulated as a routing problem 
solved by the quantum annealing, as 
described in the 'Introduction' section. 
To facilitate this approach, 
we shall introduce Ising spins 
to describe the time evolution of 
the position of the vacancy as follows: 
Let $q_{\alpha,i}$ take the value 1 when a vacancy 
is located at the site $\alpha$ at the step $i$, 
and otherwise take the value 0. 
The initial (final) condition is then 
described as $q_{S,1}=1$ ($q_{T,N}=1$). 
Under these conditions, the annealing framework 
is capable of providing optimum trajectories 
when $i=2\sim (N-1)$. 
The probability that $q_{k,N-1}=1$ corresponds 
to $P_k$ in Eq.(\ref{eq.average_cosine}). 
A trajectory is expressed by a 
spin alignment 
$\left\{q_{\alpha,i}\right\}$ 
dominated by an Ising Hamiltonian
~\cite{2017NEU, 2017SYR, 2013CRI, 2004MAR}:
\begin{eqnarray}
  \hat H_N
  &=& \sum\limits_{\alpha ,\beta } {\sum\limits_{i = 1}^{N - 1}
    {\left( {{t_{\alpha \to \beta }}\cdot {q_{a,i}}
    {q_{\beta ,i + 1}}} \right)} }
  + {\lambda _2}\sum\limits_{i = 1}^N 
  {{{\left( {\sum\limits_\alpha
          {{q_{\alpha ,i}} - 1} } \right)}^2}}
   \nonumber \\
  &&  
  + {\lambda _3}{\left( {{q_{S,1}} - 1} \right)^2}
  + {\lambda _4}{\left( {\sum\limits_{i = 2}^{N - 1} 
  {{q_{{\rm{T}},i}}} } -0\right)^2}
  + {\lambda _5}{\left( {{q_{T,N}} - 1} \right)^2} \ .
\label{hamiltonian}
\end{eqnarray}
The first term describes 
the hopping of a vacancy between sites, 
$\alpha\to\beta$.
The hopping amplitude $t_{\alpha\to\beta}$ corresponds to 
the probability of the hopping $p$, 
which scales with a temperature ($T$) dependence 
$p_{\alpha\to\beta}\sim 
\exp{\left( \Delta E_{\alpha\to\beta}/T \right)}
\sim
\exp{\left(t_{\alpha\to\beta}/T \right)}$. 
Here $\Delta E_{\alpha\to\beta}$ is the 
barrier energy for the hopping, which can be 
evaluated by {\it ab initio} calculations
~\cite{2019ICHa}. 
The amplitude $t$ is therefore 
related to $p$ by 
$t\propto \ln{p} $.
The terms with $\lambda_3$ and $\lambda_5$ 
denote the initial and final conditions 
as the constraints. 
The term with $\lambda_2$ expresses 
the condition that only one vacancy exists 
over all the sites, i.e., the
assumption that we consider a single vacancy 
contributing to the pulling-back as 
the primary contribution to $f$, ignoring 
multiple-vacancy processes as secondary.
  This assumption is reasonable except for some cases.
  Noted that most of the exceptions
  are in face-centered metallic crystals,
  where the bi-vacancy process significantly
  contributes to the self-diffusion
  when the temperature is higher than
  $2/3$ of the melting temperature ~\cite{2007MEH}.
The term with $\lambda_4$ means that 
the vacancy never exchanges its position 
with the tracer until $i=N$, 
as the problem assumes. 

\subsection{Evaluation of the correlation factor}
As a concrete example, consider a 5$\times$5 lattice
in two dimensions: 
\begin{align}
\label{fieldExam}
\left( \begin{array}{l}
{\rm{(0,0) \ (0,1) \ (0,2) \ (0,3) \ (0,4) }}\\
{\rm{(1,0) \ (1,1) \ (1,2) \ (1,3) \ (1,4) }}\\
{\rm{(2,0) \ (2,1) \ (2,2) \ (2,3) \ (2,4) }}\\
{\rm{(3,0) \ (3,1) \ (3,2) \ (3,3) \ (3,4) }}\\
{\rm{(4,0) \ (4,1) \ (4,2) \ (4,3) \ (4,4) }}
\end{array} \right) \ , 
\end{align}
where the entries in the matrix are the site indices. 
Suppose that a tracer located initially at (2,1) 
hops onto (2,2), where initially there was a vacancy. 
We then consider the process by which 
the tracer is pulled 'back' by the vacancy 
with an angle $\theta_k$ and 
probability $P_k$ 
of evaluating the average given by 
Eq.~(\ref{eq.average_cosine}). 
The process is complete when 
the vacancy coalesces with the 
tracer again ($q_{T,N}=1$). 
Contributions to the summation 
are not only from the direct 
'pulling back' ($\theta_k = \pi, N=2$) 
from (2,1) [the site 
where a new vacancy appears 
due to the tracer's hopping], 
but also from other possible sites 
at which the vacancy arrives 
after strolling for several 
time steps, as shown in 
Table~\ref{trajExam}. 
\begin{table}[htb]
  \label{trajExam}
  \caption{
    Possible trajectories for a vacancy 
    generated at (2,1) due to  
    hopping by a tracer. The vacancy 
    coalesces with the tracer again after 
    taking $(N-1)$ steps on the 2-dim. 
    lattice shown in Eq.~(\ref{fieldExam}). 
    The coalescence angle $\theta$ is 
    measured from the direction of the 
    initial hop by the tracer. 
    Each trajectory contributes to 
    the summation in Eq.~(\ref{eq.average_cosine}) 
    with weight $P_k$ corresponding 
    to the energy $E \sim \sum_{\alpha\beta} 
    {t_{\alpha\to\beta}}$. 
    For this simplified example, we set 
    $t_{\alpha\to\beta}=t$ (only between nearest 
    neighboring sites). 
    Each trajectory is identified by 
    the annealing simulation using
      the Hamiltonian
    $H_N$. 
  }
  \begin{tabular}{lccc}
    Trajectory                     & $\theta$ & Contribution & $H_N$ \\
    \hline 
    (2,1)(2,2)                     & $\pi$      & $1t$     & $H_2$ \\ 
    (2,1)(1,1)(2,1)(2,2)           & $\pi$      & $3t$     & $H_4$ \\ 
    (2,1)(2,0)(2,1)(2,2)           & $\pi$      & $3t$     & $H_4$ \\
    (2,1)(3,1)(2,1)(2,2)           & $\pi$      & $3t$     & $H_4$ \\
    (2,1)(1,1)(1,2)(2,2)           & $\pi$ /2   & $5t$     & $H_6$ \\ 
    (2,1)(3,1)(3,2)(2,2)           & 3$\pi$ /2  & $5t$     & $H_6$ \\ 
    (2,1)(1,1)(1,0)(2,0)(2,1)(2,2) & $\pi$      & $7t$     & $H_8$ \\
    ...                            &            & $  $     &       \\
    (2,1)(1,1)(1,2)(1,3)(2,3)(2,2) &  $\pi$ /2  & $7t$     & $H_8$ \\
    ...                            &          & $   $      &   
  \end{tabular}
  \label{trajExam}
\end{table}

\vspace{2mm}
Let us denote the contributions from 
trajectories obtained by the 
simulation with
  the Hamiltonian
$H_N$ as 
\begin{eqnarray}
P_k^{(N)}
=\sum_{l\in{\Omega};~{\rm trajectories}}{\pi_l}
\ , 
\label{contribution}
\end{eqnarray}
where $l$ indexes each trajectory and 
$\Omega$ is the space formed by all the 
contributing trajectories. 
Each contribution from a trajectory 
with energy $E_l^{(N)}$ would be 
expressed as 
$\pi_l\sim\exp{\left( E_l^{(N)} /T \right)}$. 
For example, in the case of $N=4$ in 
Table~\ref{trajExam}, $\pi_l\sim\exp{(3t)}\sim p^3$. 
Noticing that trajectories with different $N$ values
(numbers of steps to arrive at coalescence with 
a tracer) are mutually exclusive, 
the probability $P_k$ can be expressed 
as a sum of each exclusive contribution 
with different $N$: 
\begin{eqnarray}
{P_k} = \sum\limits_{N = 2}^{{N_{\max }}} 
{P_k^{(N)}} \ ,
 \label{eq.prob_for_N}
\end{eqnarray}
where ${P_k^{(N)}}$ is the probability 
of finding a vacancy at a neighboring site 
with ${\theta _k}$ obtained from the 
simulation with $\hat H_N$. 
${P_k^{(N)}}$ is obtained as the ratio 
of the number of trajectories with 
${\theta _k}$ divided by the total number 
of trajectories within the simulation 
using $\hat H_N$. 

\vspace{2mm}
In the procedure, 
quantum annealing computers (QACs) 
are used only to identify the 
optimal trajectories while 
the calculation of Eq.~(\ref{eq.average_cosine}) 
is made by a classical counting over 
a table like Table~\ref{trajExam}. 
To get such a table, 
the annealing simulations 
should be repeated even within a fixed 
$\hat H_N$. 
Recalling that an annealing simulation 
gives an optimal trajectory, 
enough repetition 
is required to search all the possible 
trajectories that are likely 
to be degenerate even within a $\hat H_N$. 
After all the possible trajectories 
have been tabulated, 
the calculation of 
Eq.~(\ref{eq.average_cosine}) 
by the classical counting on the table
can be attempted.
One might wonder whether it is possible to 
perform the {\it on-the-fly} evaluation of
Eq.~(\ref{eq.average_cosine}) 
during the search for optimal trajectories. 
For example, suppose that '$\theta=0$'
were obtained 5 times, possibilities. 
One might be tempted to use the frequency of 
the appearance of a particular angle for 
an 'on-the-fly' calculation of $P_k$. 
However, this cannot be justified 
at least for QAC, 
as we note later in the first 
paragraph of the 'Discussion' section. 

\section{Results and
  Discussion}
\label{sec.results}

\subsection{Verification of benchmark}
For some selected cases with 
simple lattices, it is possible to describe 
the multi-scattering processes contributing to
the correlation factor in terms of 
recursion equations, and thus to find analytical 
solutions~\cite{2007MEH}; some examples are shown 
in Table~\ref{tab.analy}. 
\begin{table}
  \begin{center}
    \caption{\label{tab.analy} 
    Correlation factors $f$, obtained 
    analytically for simple lattice model systems\cite{2007MEH}.
    }
    \begin{tabular}{c|l}
      Lattice & $f$\\
      \hline 
      Beehive & 1/3 \\
      2-Dim. Tetragonal & 0.467 \\
      2-Dim. Hexagonal  & 0.56006 \\      
      Diamond & 1/2 \\
      Simple Cubic & 0.6531 \\
      Body-Centered Cubic & 0.7272, (0.72149) \\
      Face-Centered Cubic & 0.7815 \\ 
      \hline 
    \end{tabular}
  \end{center}
\end{table}
\begin{table}
  \begin{center}
    \caption{\label{tab.N_conv}
      The convergence 
      of the correlation factors 
      evaluated by
        `(a) Quantum Annealing with D-wave (QA)',
      `(b) Simulated Quantum Annealing (SQA)',
      `(c) Classical Random Walk (CRW)', 
      and 
      `(d) Matrix Updating method (MU)', 
      depending on the system size $N$.
        ($^*$: The difference from the other methods is
        attributed to that our QA calculation could count over
        only 94.54~\% of the trajectories, because of the
        limited number of samples due to the computational cost.)
    }
    \begin{tabular}{r|cccc}
      $N_{\rm max}$ \quad & QA(a) & SQA(b) & CRW(c) & MU(d) \\
      \hline \\
      1 & -     &   -      &   -      &   -   \\
      2 & 0.600 & 0.600    & 0.600    & 0.600 \\
      3 & -     &   -      &   -      &   -   \\
      4 & 0.542 & 0.542    & 0.542    & 0.542 \\
      5 & -     &   -      &   -      &   -   \\
      6 & 0.520$^*$ & 0.519& 0.519    & 0.519 \\
      7 & -     &   -      &   -      &   -   \\
      8 & -     &   -      & 0.507    & 0.507 \\
      9 & -     &   -      &   -      &   -   \\
     10 & -     &   -      & 0.499    & 0.499 \\
     11 & -     &   -      &   -      &   -   \\
     12 & -     &   -      & 0.495    & 0.494 \\
     13 & -     &   -      &   -      &   -   \\
     14 & -     &   -      &   -      & 0.491 \\
        & $\cdots$ & $\cdots$ & $\cdots$ & $\cdots$ \\
     32 & -     &   -      &   -      & 0.477 \\
        & $\cdots$ & $\cdots$ & $\cdots$ & $\cdots$ \\
    492 & -     &   -      &   -      & 0.468 \\
        & $\cdots$ & $\cdots$ & $\cdots$ & $\cdots$ \\
    502 & -     &   -      &   -      & 0.468 \\
        & $\cdots$ & $\cdots$ & $\cdots$ & $\cdots$ \\
    \hline 
    \end{tabular}
  \end{center}
\end{table}
The values given in Table~\ref{tab.analy} 
can be used 
to test our formulation and its 
implementation. 
We are able to reproduce the value 
$f$ = 0.467\cite{1973MON}
for a two-dimensional tetragonal lattice
by our procedure, as described below. 
Note that the analytical solution is 
obtained only for a quite limited case in 
which the initial and the final positions of 
the tracer are within one step of each other, 
($T=S+1$)~\cite{2009MANb}, 
while our treatment is never limited 
by such toy-model assumption. 
The present approach is therefore capable of 
providing interesting knowledge
going beyond what can be learned by existing methods. 

\vspace{2mm}
Though '(a) Quantum annealing computers (QAC)'
are ultimately the preferred technology for counting
up trajectories to get $P_k$, 
the availability of such devices is still limited, not only 
by financial considerations, but also by 
the total number of Q-bits technically achieved. 
As explained later, current availability 
enables us to try up to $N_{\rm max}\sim 5$: 
far too few to 
verify the calibration of the two-dimensional tetragonal lattice ($f$ = 0.467\cite{1973MON}). 

\vspace{2mm}
As possible substitutes, we can list 
'(b) simulated quantum annealing~(SQA){~\cite{2002SAN, 2002MAR}}/
path integral monte carlo~(PIMC){\cite{1995CEP,2006MOR}}', 
'(c) classical random walk (CRW)', 
and '(d) matrix updating (MU)', in order of their closeness 
to (a). 
Unfortunately, for larger $N_{\rm max}$, the feasibility of (b) and
(c) proved limited.

\vspace{2mm}
For '(b) SQA', the required computational 
cost is dominated by the annealing time, i.e., the 
time to decrease the transverse magnetic field. 
To achieve the equilibrium Boltzmann distribution, 
this time  
increases with system size $N$ as $\sim\exp(N)$
{~\cite{2006MOR}}.
This limits the possible number of trajectories 
obtainable at an affordable cost, 
leading to larger error bars in Eq. (\ref{eq.prob_for_N}), 
as shown in Table~\ref{tab.N_conv}. 

\vspace{2mm}
For '(c) CRW', feasibility is assured up to 
$N_{\rm max}$=12 in the present case. 
In this method, the computational time 
is dominated by the number of stochastic 
trials. 
For a step there are $Z$ possible ways
of hopping to nearest neighboring sites 
($Z=4$ in this benchmark case). 
The total number of possibilities 
for an $N$-step trajectory amounts to 
$Z^N$, which easily becomes very large as 
$N$ increases. 

\vspace{2mm}
By using '(d) MU', we can successfully verify 
the calibration by going up to $N_{\rm max}$=500, 
as described below (Table~\ref{tab.N_conv}). 
We introduce the vacancy hopping operator
\[
\hat T 
= \sum\limits_{i,j} {{t_{ij}}\cdot a_i^\dag a_j} \ , 
\]
Consider a field described by the matrix 
\[
{F_0} = \left( {\begin{array}{*{20}{c}}
0&0&0&0&0\\
0&0&0&0&0\\
0&0&1&0&0\\
0&0&0&0&0\\
0&0&0&0&0
\end{array}} \right) \ , 
\]
where each element $(F_{0})_{i,j}$ corresponds 
to the location of a hopping site. 
The value '1' in $F_0$ indicates the 
(initial) location of a vacancy, 
whereas '0' means the vacancy is absent.
We update the field at step $K$ to $F_K$, by 
\begin{eqnarray}
{F_K} = \hat T\cdot{F_{K - 1}} \ .
\label{thop}
\end{eqnarray}
In the present case (two-dimensional tetragonal lattice), 
we assume $t_{ij}$ is isotropic and 
only connects
between the nearest neighboring sites. This  
drives the field matrix to 
\[
{\left( {{F_K}} \right)_{i,j}} 
= {\left( {{F_{K - 1}}} \right)_{i - 1,j}} 
+ {\left( {{F_{K - 1}}} \right)_{i + 1,j}} 
+ {\left( {{F_{K - 1}}} \right)_{i,j + 1}} 
+ {\left( {{F_{K - 1}}} \right)_{i,j - 1}} \ . 
\]
The constraint that the vacancy not  
coalesce with the tracer until the given  
final step $N$ can be expressed as 
${\left( {{F_K}} \right)_{i',j'}}=0$
for $K < N$ 
where $\left(i',j'\right)$
is the location of 
the tracer site. 
After updating the field matrix 
until step $N$, each matrix element 
shows how many trajectories being possible 
to give a vacancy at that site 
after $N$ steps, from which we can 
evaluate $P_k$ and thus $f$. 
As shown in Table~\ref{tab.N_conv}, 
$f$ falls as $N$ increases. 
It is at 0.468 when $N = 500$, 
and the rate of decline has 
become very small. Thus, it appears 
to be asymptotically approaching 
the value from the analytical solution, 0.467.

\vspace{2mm}
The feasibility of '(a) Quantum annealing computers (QAC)' 
is determined in large part by the available number of 
Q-bits, $N_{\rm Qbit}^{(\rm available)}$, 
currently $2048$,\cite{2020DWAc}. 
The {\it required} number of Q-bits scales in the present case 
as the product of $N_{\rm max}$ and the 
size of the lattice ($M\times M$ in the two-dimensional 
case; $5\times 5$ in the example). 
Therefore, the maximum possible 
$N_{\rm max}^{(\rm possible)}$ may
be estimated as 81 (= 2048/25);
for a user with a
practical budget situation, it is probably closer to five.
We note however that the computational 
limitation of being directly and linearly
proportional to $N_{\rm Qbit}^{(\rm available)}$
still renders (a) 
more promising than other methods like (b) and (c).

\vspace{2mm}
For '(a)~QAC', we used D-Wave \cite{2020DWAa} applied 
to $(N_{\rm max}+1) \times (N_{\rm max}+1)$ lattice size 
for $N_{\rm max}=2,4,6\cdots$ in order. Since implemented
topologies of Q-bits interconnections (chimera graph) 
are not capable in general to describe Ising spin couplings 
as it is in the Hamiltonian, some of the couplings 
(spins directly couple with each other in the Hamiltonian,
say $J_{12}\sigma_1\sigma_2$) are equivalently realized 
by the synchronized Q-bits pairs 
({\it i.e.}, $\sigma_1$---$\sigma_2$ in the Hamiltonian 
is realized as $\sigma_1$---$\tau_1$...$\tau_2$---$\sigma_2$, 
where $\tau_{1}$ and $\tau_{2}$ 
are distant but synchronized). \cite{2019FOS}
The technique costs the number of Q-bits than that 
of pure required one in the model Hamiltonian. 
Even using 2,000 Q-bits, 
we could embed our problem only upto $N_{\rm max}=2$ 
on the D-wave using the technique. 
In such a case, 
we can use the 'Hybrid solver'
to resolve the problem. \cite{2020DWAb}
The solver itself works on a classical 
computer, decomposing the original size problem 
into a set of smaller chimeric graphs 
those are possible to be handled by D-wave. 
The set of results by D-wave is 
then post-processed by the solver 
to get the answer of the original problem 
on the classical computer.
  By using the solver, we have confirmed that proper
  trajectories are obtained upto, at least, $N_{\rm max}=12$.
  However, to get the correlation factor finally, 
  we have to count over all the trajectories,
  for which we could achieve upto $N_{\rm max}=6$
  due to the D-wave resource limitation.
  For $N_{\rm max}$=2, 4, and 6, we sampled 1, 15,
  and 240 solutions, covering 100~\%, 100~\%, and
  94.54 \% of the trajectories, respectively. 
All the above limitations are, however, 
coming purely from the technical/implementational 
aspect of Quantum Annealing machines. 
It is straightforward for us to make the 
limitations ahead assisted by 
the intensive developments on the 
implementations such as the 
increasing $N_{\rm Qbit}^{\rm (available)}$, 
improved topologies of chimera graph {\it etc.}
(\textit{e.g.}, pegasus graph \cite{2020BOO}).
We note that the intrinsic computational cost for 
the trajectory sampling is just several $\mu$sec. 
as we confirmed. 

\subsection{Discussions}
In the procedure explained above, 
it is assumed that all the degenerate 
ground state spin configurations 
(i.e., the optimal trajectories) 
can be found after a sufficiently large 
(but finite) numbers of trials 
of the annealing simulation. 
We should note, however, that
there seems to be no firm theoretical 
basis for this assumption. 
In SQA, by contrast, 
it is guaranteed that all 
degenerate states will be realized 
under the Boltzmann distribution 
if the transverse 
magnetic field is decreased 
by the correct procedure{~\cite{2006MOR}}. 
For QAC, we could not find such 
a clear foundation, but the literature 
seems to support our assumption. 
It has been reported that a D-Wave machine 
can realize the optimal states 
dominated by the Boltzmann distribution 
under an ideal operation{~\cite{2010BIA}}.
There is also a report that, in the setting 
of quadratic unconstrained binary optimization, 
Gaussian noise intentionally 
added on the coefficients {\it improves} 
the reproducibility of 
simulations.{~\cite{2019FOS}}
If the unsatisfactory reproducibility 
was due to the 'bias in the frequency 
to get equivalent degenerate solutions',  
then the {\it improvement} seems to 
correspond to a hopeful procedure 
to ensure our assumption here. 
It is interesting to estimate how much error 
will occur in the correlation factor $f$ 
when some degenerate trajectories 
are missing from the count. 
Larger multiplicities in 
the degeneracies occur in 
the large $N$ region, 
for which MU~($N_{\rm max} = 501$) 
is currently the only means of access. 
We intentionally dropped off 
some of the degenerate trajectories 
randomly (at most 10\%). 
The bias in the estimated $f$ was then 
found to be $\sim 0.4$\%. 
             
\vspace{2mm}
Given the present value of 
$N_{\rm Qbit}^{(\rm available)}$, 
MU is still superior to QAC 
. 
It is therefore important to 
discuss what 
restricts further scalability 
of MU, and what will make QAC inherently 
superior when  
$N_{\rm Qbit}^{(\rm available)}$ 
is larger. 
In the space $\Omega$ of all 
trajectories (mentioned in
in Eq.~(\ref{contribution})), 
the weight, 
$\exp{\left(-\beta E_l^{(N)}\right)}$, 
dominates only for 
those trajectories with the most 
stable energy $E_0^{(N)}$ at 
lower temperature. 
Denoting the space formed by 
such (possibly degenerate) trajectories 
with the lowest energy as $\mathcal{A}\subset \Omega$,
then 
\[P_k^{(N)}\sim\sum_{l\in{\mathcal{A}}}{\pi_l} \ , \]
for the temperature range. 
The advantage of QAC in 
optimization problems in general 
is its 
quite efficient ability to extract 
$\mathcal{A}$ from $\Omega$. 
MU, on the other hand, is a 
scheme which surveys
all the elements of $\Omega$, 
since it accumulates the number 
of visits $N_{\rm visits}$ 
by the vacancy to every lattice site. 
When the system size is very large, 
$|\mathcal{A}| \ll |\Omega|$, 
and hence QAC will perform 
more efficiently than MU 
in evaluating $P_k^{(N)}$. 
From this viewpoint, 
the present benchmark, 
the two-dimensional tetragonal lattice, 
would be highly 
inappropriate for showing 
the superiority of QAC for the 
following reason:
In the simplified case 
($t_{\alpha\to\beta}=t$), 
all the trajectories having 
the same $N$ have the same 
energy and are elements of $\mathcal{A}$. 
Hence $\mathcal{A}= \Omega$ 
and the advantage of QAC 
disappears. 

\vspace{2mm}
MU can easily be generalized to 
higher dimensional lattices with 
general shapes and with anisotropic 
hopping. 
The temperature dependence of 
the hopping can be parameterized 
via the factor 
$\exp{\left(-\beta E_l^{(N)}\right)}$, 
and then the scheme would be 
useful for analyzing temperature-depending 
diffusion (as would QAC). 
In the case of the two-dimensional 
tetragonal lattice, however, 
the success of MU with 
$N_{\rm max}\sim 500$ 
is in fact just 
a lucky accident due to the presence of an especially 
efficient data structure valid only 
for this case. 
The factor dominating $N_{\rm max}$ 
in MU comes from the upper limit 
of the largest possible exponent 
of $N_{\rm visits}$, represented by 
various numeric data types. 
It increases explosively 
in factorial manner as $N$ increases, 
and (using integer type) easily 
overflows. 
In the present work, we use instead 
the double precision type
with mantissa/exponent representation, 
and find the upper limit of the exponent 
corresponds to $N_{\rm max}\sim 500$ 
even using the simplest possible data 
structure to store $N_{\rm visits}$. 
When we try more general cases, 
such as three-dimensional lattices, we cannot 
use such a simple data structure but 
instead must use 'struct' type to store 
$N_{\rm visits}$, leading to a
much reduced $N_{\rm visits}\sim 20$ 
(for the three-dimensional cubic lattice). 

\vspace{2mm}
The difficulty of 
accommodating $N_{\rm visits}$ in a 
practical size of data storage 
comes from the fact that 
MU has to treat all the 
trajectories in $\Omega$. 
QAC, on the other hand, has no such 
inherent problem, because it only 
deals with $\mathcal{A}$. 
The method is then potentially 
feasible 
in the future when 
$N_{\rm Qbit}^{\rm available}$ 
increases. 
  
\section{Conclusion}
\label{sec.conc}
We developed a framework to evaluate the
correlation factor, a key quantity used 
to derive the macroscopic diffusion coefficient 
for ions in solid materials. 
The coefficient describes the process 
by which a vacancy attracts back a tracer 
even after repeated scattering events. 
Direct counting of the possible processes 
is not feasible with conventional computational 
tools, so the coefficient has previously only been 
evaluated in limited model 
cases where simple assumptions 
allowing the process to be described 
in terms of recursion formulae can be justified. 
This has hampered the utilization of microscopic 
information obtained by {\it ab initio} 
approaches (vacancy rate, 
formation energy for a defect, 
energy barrier to hopping, etc.) 
in macroscopic calculations. 
By using our framework, 
we verified as a calibration that direct counting
reliably reproduces the results obtained 
previously by the recursion model. 
The framework promises to be especially valuable 
when implemented on quantum computers 
with the increased number of 
available Q-bits made possible by recent 
technological advances. 
The applicability of the direct counting 
approach is never 
restricted to special cases, so we can investigate how 
the diffusion coefficient is affected by 
nano-level tuning of materials and other factors 
evaluated by {\it ab initio}
calculations, factors not previously 
considered applicable to practical ionic hopping 
networks in realistic materials. 

\section{Acknowledgments}
The computation in this work has been performed using the facilities
of the Research Center for Advanced Computing Infrastructure (RCACI) at JAIST.
T.I. is grateful for financial support from Grant-in-Aid for
JSPS Research Fellow (18J12653).
K.H. is grateful for financial support from a KAKENHI grant (JP17K17762),
a Grant-in-Aid for Scientific Research on Innovative
Areas ``Mixed Anion`` project (JP16H06439) from MEXT, 
PRESTO (JPMJPR16NA) and the Materials research by Information Integration
Initiative (MI$^2$I) project 
of the Support Program for Starting Up Innovation Hub from Japan Science and
Technology Agency (JST). 
R.M. is grateful for financial support from MEXT-KAKENHI (17H05478 and 16KK0097), 
from Toyota Motor Corporation, from I-O DATA Foundation, 
and from the Air Force Office of Scientific Research (AFOSR-AOARD/FA2386-17-1-4049).
R.M. and K.H. are also grateful to financial support from
MEXT-FLAGSHIP2020 (hp170269, hp170220).

\bibliographystyle{apsrev4-1}
\bibliography{references}
\end{document}